\documentclass[pra,showpacs,twocolumn,amsmath]{revtex4}
\usepackage{amssymb}
\usepackage{graphicx}
\usepackage{amsmath}
\usepackage{amsbsy}
\usepackage{amsthm}
\usepackage[ansinew]{inputenc} 
\usepackage{bbm}
\usepackage{bm}
\usepackage{epsfig}
\usepackage{lscape}
\usepackage{bbold}
\newcommand{\bra}[1]{\ensuremath{\left\langle #1\right|}}
\newcommand{\ket}[1]{\ensuremath{\left|#1\right\rangle}}
\newcommand{\beq}{\begin{eqnarray}}
\newcommand{\eeq}{\end{eqnarray}}

\newcommand{\tr}{\textnormal{Tr}}

\newcommand{\scal}[2]{\bra{#1} \, #2 \, \rangle}

\newcommand{\adj}[1]{#1^{\dagger}}

\begin{document}
\title{
Partial Multipartite Entanglement in the Matrix Product State Formalism}
\author{Andreas Gabriel}
\author{Valentin Murg}
\author{Beatrix C. Hiesmayr}
\email{Beatrix.Hiesmayr@univie.ac.at}

\affiliation{University of Vienna, Faculty of Physics, Boltzmanngasse 5, 1090 Vienna, Austria}

\begin{abstract}
We present a method to apply the well-known matrix product state (MPS) formalism to partially separable states in solid state systems. The computational effort of our method is equal to the effort of the standard density matrix renormalisation group
(DMRG) algorithm. Consequently, it is applicable to all usually considered condensed matter systems where the DMRG algorithm is successful. We also show in exemplary cases, that polymerisation properties of ground states are closely connected to properties of partial separability, even if the ground state itself is not partially separable.\end{abstract}


\maketitle

\section{Introduction}
Academically speaking, entanglement is one of the most fundamentally nonclassical features of quantum mechanics and as such is highly important to the foundations of modern physics, e.g., in high energy physics \cite{foundations2,foundations1}. In a more practical view, entanglement is most relevant in two distinct respects: technological applications (such as e.g. quantum cryptography \cite{crypto1,crypto2} or future quantum computers \cite{comp1,comp2,comp3}), and its --still widely unknown-- role in Nature. The latter has attracted a high degree of attention in the past years - it has e.g. been speculated, that entanglement might be involved in the extraordinary efficiency of photosynthetic light harvesting complexes \cite{photo} or geographical orientation of birds \cite{birds}. It has even been suggested, that entanglement might play a significant role in processes as macroscopic as evolution as a whole \cite{evolution}.\\
While these examples are still more speculation than scientific reality, it has been conclusively shown that entanglement is indeed closely related to macroscopic properties of more simple systems in solid state physics, such as spin chains or crystal lattices \cite{manybodyent}. For example, frustration \cite{manybodyfrust} of or certain phase transitions \cite{manybodyphase} in such systems are known to be connected to their entanglement properties.\\
While generic entanglement has been studied in the context of condensed matter since the turn of the century quite extensively (see e.g. \cite{mbe1,mbe2,mbe3,mbe4,mbe5,mbe6,mbe7,mbe8,mbe9}), only very few works address questions concerning partial (in-)separability or genuine multipartite entanglement~\cite{manybodyps,gabrielhiesmayrmanybody,GiampaoloHiesmayr,Sanpera,frustratedspingme}. In Ref.~\cite{manybodyps} partial (in-)separability was firstly mentioned but not elaborated. The authors of Ref.~\cite{gabrielhiesmayrmanybody} introduced macroscopic observables (such as energy) that are capable of detecting genuine multipartite entanglement (GME) or partial inseparability via comparing the minimal energy of the ground state to the minimal energy when optimised over the set of partial separable states. The obtained GME-gap- or partial-entanglement-gap energy works well in a wide variety of systems, the only requirement being that the ground states are not separable. However, for large numbers of particles this approach becomes computationally highly demanding. In Ref.~\cite{GiampaoloHiesmayr} the authors investigated the $XY$-model in an external magnetic field where analytical solutions of the ground states are known, allowing direct application of criteria detecting genuine multipartite entanglement. In another recent publication~\cite{Sanpera} the genuine tripartite entanglement of the anisotropic $XXZ$ spin model was analysed by applying proper constructed entanglement witnesses. In Ref.~\cite{frustratedspingme} the behavior of genuine multipartite entanglement of paradigmatic frustrated quantum spin systems was investigated by using a geometric measure.

The aim of this work is to contribute to this field by adapting existing methods for investigations of solid state systems - namely the matrix product state (MPS) formalism - to problems of partial separability in these systems. The article is organised as follows: in the upcoming section, basic definitions on entanglement, the MPS formalism and solid state systems will be reviewed. After that, we can present our main result, which is a method to use the MPS formalism to investigate partial separability in solid state systems. We then present and discuss illustrative examples before the paper is concluded.\\

\section{Definitions}
One of the most common problems in many body physics is finding the ground state energy for a given Hamiltonian $\mathcal{H}$. This task can be approached by means of the MPS formalism, in which a general state $\ket{\Psi}$ is given in the MPS basis
\beq \ket{\Psi} = \sum_{s_1,s_2,...,s_n} \ket{s_1, s_2, ..., s_n} \tr(A^1_{s_1}\cdot A^2_{s_2}\cdot ...\cdot A^n_{s_n}) \eeq
where each sum runs over the respective subsystem (i.e. from $0$ to $d-1$ for a $d$-dimensional subsystem) and the $A^i_{s_i}$ are $D\times D$ dimensional matrices. Here, $D$ is the bond dimension, i.e. the virtual dimension of the subsystems (for more details on the MPS formalism basics, see e.g. Ref.\cite{mps}).

Using this representation of $\ket{\Psi}$, the expectation value $\bra{\Psi}\mathcal{H}\ket{\Psi}$ can be computed in a very efficient manner using the density matrix renormalisation group (DMRG) algorithm \cite{dmrg}. In particular, the computational complexity of such optimisations is linear in the number $n$ of subsystems, as each $A^i_{s_i}$ can be optimised essentially individually.

Despite its simplicity the concept of partial separability turned out to be a meaningful one in respect to a classification scheme for multipartite entanglement~\cite{revistedpartialseparability}.
Moreover, most complications only arise for mixed states, which are of no concern to the issues addressed in this work. Consequently, we will only discuss the much more simple and illustrative pure state case.

An $n$-partite quantum state $\ket{\Psi}$ is called $k$-separable, iff it can be written as a nontrivial $k$-fold product
\beq \ket{\Psi} = \ket{\psi_1}\otimes\ket{\psi_2}\otimes...\otimes\ket{\psi_k} \eeq
with $1 \leq k \leq n$, where each of the factors $\ket{\psi_i}$ is a state of one or several subsystems. If a state is separable under a given $k$-partition (as opposed to an unknown partition) $\gamma = \{\gamma_1,\gamma_2,...,\gamma_k\}$ (with $\bigcup_i \gamma_i = \{1,2,...,n\}$ and $\gamma_i\cap\gamma_j = \{\}$ for $i\neq j$) it is called $\gamma_k$-separable. A state which does not factorise at all (i.e. which is not 2-separable) is called genuinely multipartite entangled.\\
For example, the 4-partite state
\beq \ket{\Psi}=\frac{1}{2}\left(\ket{0000}+\ket{1001}+\ket{0010}+\ket{1011}\right) \eeq
is 3-separable, as it can be written as $\ket{\Psi}=\ket{\Phi^+}_{1,4}\otimes\ket{0}_2\otimes\ket{+}_3$. Consequently, it is also biseparable (as any two of the three factors can be formally combined into one). It is, however, not 4-separable, since there is no basis in which the first and fourth subsystems factorise. Note that any form of partial separability that is not full separability (i.e. $n$-separability) is equivalent to partial entanglement.\\

\section{Using the MPS Formalism for Partial Separability}
Since the MPS formalism relies on the state in question being given in the MPS basis, a straightforward implementation of partial separability is impossible (as $k$-separability is defined via the freedom of choice of local bases for state vectors). However, the problem of minimising the energy $\bra{\Psi}\mathcal{H}\ket{\Psi}$ over states with given separability properties only (instead of all states), can be approached by imposing these separability constraints on the MPS structure before optimising.

\begin{figure}[ht!]\
\centering\includegraphics[width=8cm]{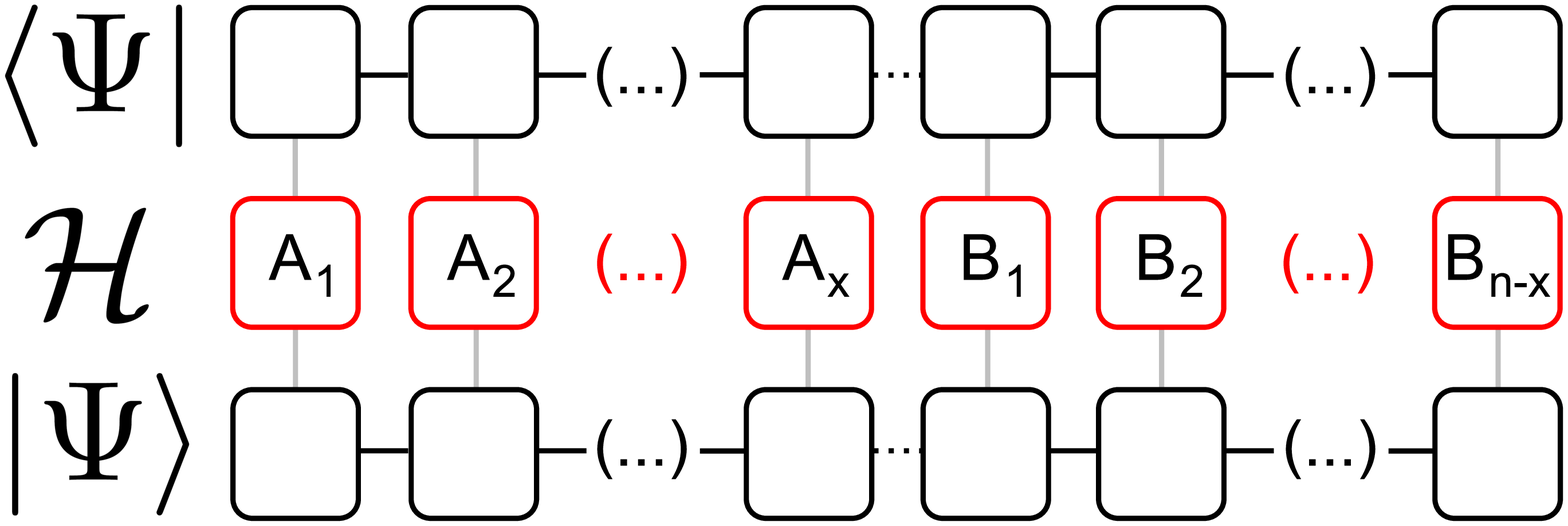}
\caption{Illustration of the scalar product $\bra{\Psi}\mathcal{H}\ket{\Psi}$ corresponding to the energy expectation value for a 2-separable state $\ket{\Psi}$. Here, $\ket{\Psi}$ is separable with respect to the partition $\gamma=\{A_1,A_2,...A_x | B_1,B_2,...B_{n-x}\}$. The red squares correspond to the Hamiltonian's sites (reordered according to the partition $\gamma$), the black squares correspond to the MPS's sites. The horizontal black lines connecting the MPS sites correspond to the virtual bonds of dimension $D$, the two dotted lines depict 1-dimensional bonds which provide the desired separability properties and the vertical gray lines correspond to the physical $d$-dimensional indices.}\label{fig_mps}\end{figure}

For example, an $n$-partite state separable under the bipartition $\{1,2,...,x | x+1,x+2,...n\}$ (for some $1 \leq x < n$) can be realised by setting the bond dimension $D$ between sites $x$ and $x+1$ to 1, thus forcing the state into two mutually separable and independant blocks (note that in the case of periodic boundary conditions, the bond dimension between sites 1 and $n$ also has to be reduced to 1 in this case). In order to also be able to implement non-compact partitions (i.e. partitions where the parts do not form blocks, but overlap, such as e.g. $\{1,3,5,... | 2,4,6,...\}$), the enumeration of sites can be permuted accordingly in the Hamiltonian (as illustrated in Fig.~\ref{fig_mps})
.\\
Using this approach, the set of states accessible to a MPS can be effectively limited to all states separable under an arbitrary partition. The computational complexity is reduced in contrast to the case with additional constraints.

\section{Examples}
In order to investigate the connection between polymerisation and partial separability, let us introduce the partition step parameter $p$, which uniquely defines the partitions we want to investigate. Specifically, these partitions consist of blocks of $p$ sites, alternatingly allocated to two parts of a bipartition. In other words the spin chain is split into $\frac{n}{p}$ blocks of $p$ subsystems each, where the first, third, and all other odd-numbered blocks are separable from all even-numbered blocks. In order to better simulate the thermodynamical limit (i.e. many particles), there should always be equally many even- and odd-numbered blocks, which implies that $n$ should always be an even multiple of $p$.\\

\subsection{Dimerised Heisenberg Model}

\begin{figure}
\centering\includegraphics[width=8cm]{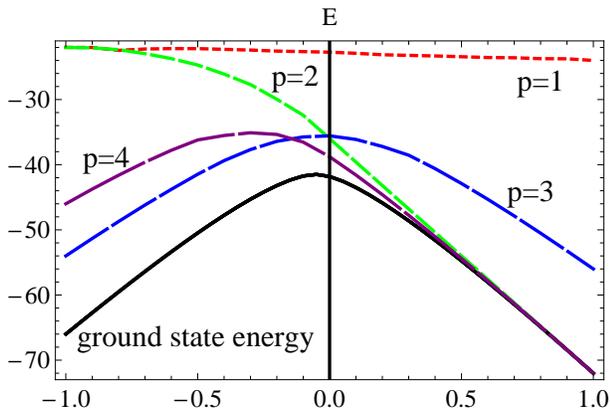}
\caption{Minimal energies $E$ for states with different separability properties in the dimerised Heisenberg model of $n=24$ particles varied over $\delta$ is shown. The black line depicts the ground state energy, while the coloured dashed lines correspond to minimal energies for biseparable states with different $p\in \{1,2,3,4\}$. The slight shift in the peak of the ground state energy curve as well as the slight change of the $p=1$ curve with $\delta$ are due to the open boundary conditions used in these calculations.}
\label{fig_dimH}
\end{figure}

As a first example, we consider the dimerised Heisenberg model Hamiltonian
\beq \mathcal{H} = \sum_i (1-(-1)^i\delta) \vec{s}_i\cdot\vec{s}_{i+1} \eeq
of a chain of $n$ spin-$\frac{1}{2}$ particles. For $\delta\neq0$, the ground state of this Hamiltonian is known to exhibit dimerisation properties \cite{dimerisingheisenberg,chitra95}. The different minimal energies for partitions corresponding to different values of $p$ are plotted in Fig.~\ref{fig_dimH}.
As expected, for $\delta \rightarrow1$ (i.e. complete dimerisation) the energies for $p=2$ and $p=4$ coincide with the ground state energy. Moreover, in the entire range $0 < \delta \leq 1$ these energies are very close to the ground state energy, as each block of 2 or 4 sites contains 1 or 2 pairs of dimerised particles, respectively, and is thus optimally chosen (while e.g. for $p=3$ there is one 'unallocated' site in each block). Since for $\delta<0$ the favoured partitions are different (all site numbers are shifted by one), these energies do not indicate polymerisation in this area. Whether such behaviour of energies can be used as a sufficient criterion for polymerisation is not entirely clear, however, it does appear to be a necessary condition.\\

\subsection{Bilinear Biquadratic Model}

An example with particularly interesting dimerisation properties is the bilinear biquadratic (BLBQ) model Hamiltonian
\beq \mathcal{H} = \sum_{\left\langle i,j\right\rangle} \left(\vec{S}_i\cdot\vec{S}_j + \alpha (\vec{S}_i\cdot\vec{S}_j)^2\right) \eeq
of a chain of spin-1 particles, where the sum runs over all nearest-neighbor pairs.
Such quantum magnetic systems can be realised e.g. by
spinor atoms in an optical lattice, such as $^{23}\textrm{Na}$ with a total moment $S=1$.
By confining the atoms to an optical lattice, there
are two scattering channels for identical atoms with total spin
$S=0,2$ which can be mapped to an effective BLBQ spin interaction~\cite{ripoll04}.
The Hamiltonian's ground state is known to possess a rich structure: it shows a dimer structure for $\alpha < -1$,  a trimer structure for $\alpha > 1$ and a Haldane phase inbetween~\cite{blbq}. The model is Bethe-Ansatz~\cite{Bethe} solvable at the points $\alpha = \pm 1$ and simplifies to the AKLT-model~\cite{AKLTmodel} for $\alpha=1/3$.

\begin{figure*}
\centering\includegraphics[width=14cm]{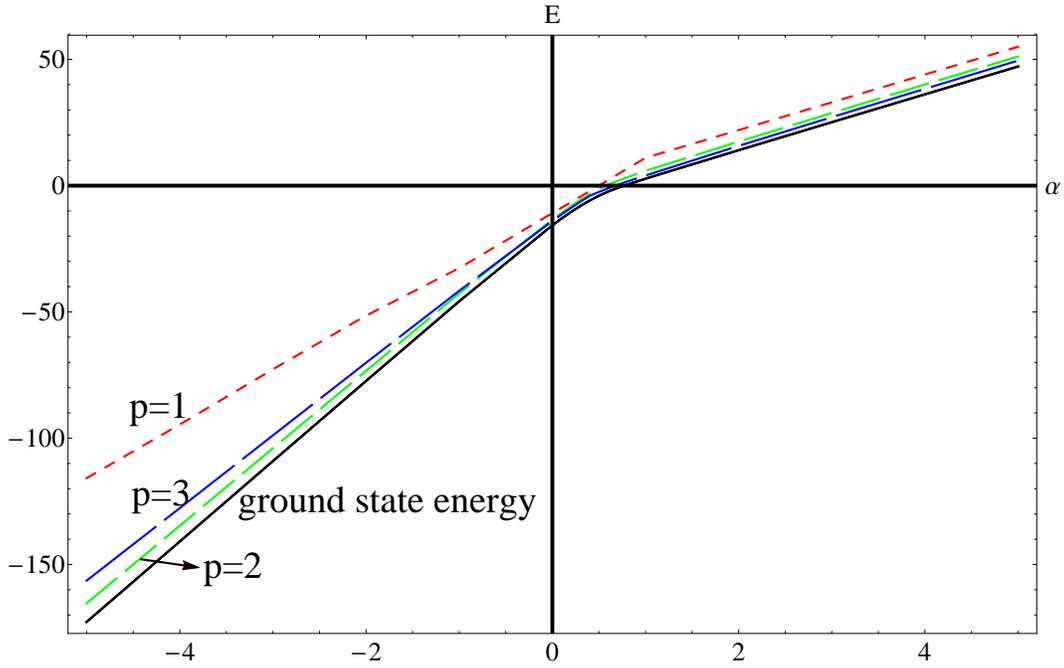}\caption{Minimal energies of biseparable states for the BLBQ-Hamiltonian for $n=12$ spin-$1$-particles. The black line depicts the ground state energy, while the coloured dashed lines correspond to minimal energies for biseparable states with different values $p=1,2,3$.}
\label{fig_BLBQ}
\end{figure*}

The structure of the ground state is reflected in the minimal energies for biseparable energies under the respective partitions as shown in Fig.~\ref{fig_BLBQ}. As expected, for $\alpha <-1$ the partition $p=2$ approaches the ground state energy more closely than the other partitions. For $\alpha>1$ the partition $p=3$ is closest to the ground state energy. The biggest deviation from the ground state energy is obtained for $p=1$, where all sites are separated from their nearest neighbors (the biggest relative difference is achieved for negative values of $\alpha$). The behaviour of the minimal energy function in the interval $\alpha\in[0,1]$ is quite similar for all partitions into blocks and thus does not conclusively reveal any deeper structures.

\section{Conclusion}
We presented a simple way to use the DMRG-algorithm of MPSs for partially separable states. This is done by imposing a structure of mutually separable blocks on the MPS and reordering the subsystems in the Hamiltonian according the the desired partition. Compared to usual applications of the DMRG-algorithm, our method does not increase the computational effort, since no external constraints are necessary (in fact, the computational effort is even slightly reduced due to the imposed 1-dimensional bonds).

Applying our method to systems with known polymerisation features, we could find a direct connection between the minimal energies attainable for states partially separable under different partitions and these polymerisation-properties, even in cases where the ground state itself is not separable at all.

In future applications, this method can also directly be applied to
multidimensional spin lattices using Projected Entangled Pair States (PEPS)~\cite{mps} instead of MPS, thus allowing for a deeper investigation of
partial entanglement in more complex and realistic systems. This might reveal yet unknown connections between entanglement and other properties, such as various phase transitions.
\\
\\
\textbf{Acknowledgements:}\\
AG and BCH gratefully acknowledge the Austrian Research Fund (FWF) projects FWF-P21947N16 and FWF-P23627N16, respectively.

\appendix

\section{The Algorithm}

For completeness, we review the variational calculus with MPS as it is used in our paper.
The goal is to determine the MPS
\begin{displaymath}
\ket{\Psi} = \sum_{s_1,s_2,...,s_n} \ket{s_1, s_2, ..., s_n} \tr(A^1_{s_1}\cdot A^2_{s_2}\cdot ...\cdot A^n_{s_n})
\end{displaymath}
with given dimensions $D_i \times D_{i+1}$ of the matrices $A^i_s$ which minimises the energy
\begin{equation} \label{eqn:hexpect1}
E = \frac{\bra{\Psi} H \ket{\Psi}}{\scal{\Psi}{\Psi}}.
\end{equation}
$D_i$ is set equal to $1$ or $D$, depending on the chosen partition.
If the sites in one partition are not numbered consecutively, the site indices
in the Hamiltonian are permuted accordingly, thus making the Hamiltonian non-local.
Following Ref.~\cite{verstraeteporras04} the idea is to iteratively optimise the tensors $A^i$ one by one while fixing all the other ones until convergence is reached. The crucial observation is the fact that the exact energy of $\ket{\Psi}$ (and also its normalisation) is a quadratic function of the components of the tensor $A_i$ associated with \emph{one} lattice site~$i$. Because of this, the optimal parameters $A_i$ can simply be found by solving a generalised eigenvalue problem.

The challenge that remains is to calculate the matrix--pair for which the generalised eigenvalues and eigenvectors shall be obtained. In principle, this is done by contracting all indices in the expressions $\bra{\Psi} H \ket{\Psi}$ and $\scal{\Psi}{\Psi}$ except those connecting to $A^i$. By interpreting the tensor $A^i$ as a $d D_i D_{i+1}$--dimensional vector $\boldsymbol{A}^i$, these expressions can be written as
\begin{eqnarray}
\label{eqn:expect}
\bra{\Psi} H \ket{\Psi} & = & \adj{(\boldsymbol{A}^i)} \mathcal{H}^i \boldsymbol{A}^i \\
\label{eqn:norm}
\scal{\Psi}{\Psi} & = & \adj{(\boldsymbol{A}^i)}  \mathcal{N}^i \boldsymbol{A}^i.
\end{eqnarray}
Thus, the minimum of the energy is attained by the generalised eigenvector $\boldsymbol{A}^i$ of the matrix--pair $(\mathcal{H}^i,\mathcal{N}^i)$ to the minimal eigenvalue~$\mu$:
\begin{displaymath}
\mathcal{H}^i \boldsymbol{A}^i = \mu  \mathcal{N}^i \boldsymbol{A}^i
\end{displaymath}

The matrix $\mathcal{H}^i$ can be efficiently evaluated and $\mathcal{N}^i$ can be set equal to the identity by meeting the orthonormalisation conditions $\sum_s (A^j_s)^{\dagger} A^j_s = \mathbb{1}$ for $j<i$ and $\sum_s A^j_s (A^j_s)^{\dagger} = \mathbb{1}$ for $j>i$ \cite{mps}. In this way, the optimal $A^i$ can be determined, and one can proceed with the next site, iterating the procedure until convergence.

\end{document}